\documentclass[preprint2]{aastex}

\shorttitle{Nuclear waiting points and double peaked bursts}
\shortauthors{Fisker et al.}
\begin{document}
\title{The nuclear reaction waiting points, ${}^{22}\textrm{Mg}$, ${}^{26}\textrm{Si}$, ${}^{30}\textrm{S}$, and ${}^{34}\textrm{Ar}$, and bolometrically double peaked type I X-ray bursts}
\author{Jacob Lund Fisker}
\email{fisker@quasar.physik.unibas.ch}
\author{Friedrich-Karl Thielemann}
\affil{Department of Physics and Astronomy, 
        Klingelbergstrasse 82, 4056 Basel,
	Switzerland}

\begin{abstract}
Type I X-ray bursts with a double peak in the bolometric luminosity have been observed from several sources. 
The separation between the two peaks are on the order of a few seconds. 
We propose a nuclear waiting point impedance in the thermonuclear reaction flow to explain these observations. 
Nuclear structure information suggests the potential waiting points: ${}^{26}\textrm{Mg}$, ${}^{26}\textrm{Si}$, ${}^{30}\textrm{S}$, and ${}^{34}\textrm{Ar}$, which arise in conditions, where a further reaction flow has to await a $\beta^+$-decay, because the $(\alpha,p)$-reaction is too weak to overcome the target Coulomb-barrier and the $(p,\gamma)$-reaction is quenched by photo-disintegration at the burst temperature. 
The conclusion is that the effects of the experimentally unknown ${}^{30}\textrm{S}(\alpha,p){}^{33}\textrm{Cl}$ and ${}^{34}\textrm{Ar}(\alpha,p){}^{37}\textrm{K}$ might be directly visible in the observation of X-ray burst light curves.
\end{abstract}

\keywords{X-rays: bursts --- nuclear reactions, nucleosynthesis, abundances --- stars: neutron}

\section{Introduction}
A type I X-ray burst results when a thermal instability in the electron degenerate atmosphere of a neutron star causes a thermonuclear runaway in the accreted matter.
The sudden release of nuclear energy heats the atmosphere rapidly, increasing the luminosity a few orders of magnitude above the persistent luminosity within a few seconds, after which the luminosity decreases exponentially as the atmosphere cools again.
While most of these bursts show a single peak in the luminosity, some burst observations have shown a double peaked structure in the bolometric light curve (\cite{Sztajno85,Paradijs86,Penninx89,Seon00} and \cite{Smale01}). 
Several theories have been proposed to explain such double peaks: Heat transport impedance in a two-zone model \citep{Regev84}, interactions with the accretion disc (\cite{Melia92}), a result of a $\textrm{H}$ flash developing into a combined H/He flash \citep{Hanawa84}, shear instabilities \citep{Fujimoto88} or increased proton capture on heavier nuclei \citep{Ayasli82}. 

In this letter we propose that an interaction between the helium flash shell and a waiting point impedance in the $rp$-process of the shells above it explains burst shape structure spanning over a time-scale of a few seconds. 
Since a self-consistent one-dimensional model attains a limit cycle equilibrium with repeated self-similar bursts, comparisson with observations will also place restrictions on the accretion rate.

\section{The first burst peak: A helium flash}\label{}
Observationally burst light curves differ greatly in shape but their understanding are theoretically divided into three cases, depending on whether the ignition and subsequent runaway occurs in a hydrogen rich shell, a helium rich shell, or in a mixed H/He shell \citep{Fujimoto81}. 
Burst sources are situated in low mass binary systems and generally accrete a mixture of hydrogen and helium. 
Since the atmosphere of the neutron star has a temperature of $T_9 \sim 0.1-0.2$, the $\beta$-limited hot-CNO cycle \citep{Wallace81} converts the hydrogen into helium. 

For a constant accreting rate\footnote{The Eddington critical accretion rate equals $1.1\cdot 10^{18} \textrm{g}/\textrm{s}$ ($1.7\cdot 10^{-8} M_\odot/\textrm{yr}$) and assumes temperature independent Thompson scattering in an atmosphere with a solar composition.} of $\dot{M}=0.045 \dot{M}_{Ed.}$ it requires about half a day before the continuous accretion process has pressed the matter down to a depth, where the helium which has now been formed in significant quantities becomes thermally unstable.
The accumulated helium at the base burns completely within tenths of a second and establishes a temperature gradient which is sufficiently large to allow a highly efficient convective heat transport to the surface. Observationally this results in a rise time of the surface luminosity which is less than a second \citep{Wallace82} and it is the cause of the first peak.

\section{The second burst peak: A nuclear waiting point impedance}\label{impedance}
The helium burning region radiates and convects heat upwards so that hydrogen which has not yet had the time to be fully converted into helium ignites as well and burns via the $rp$-process \citep{Wallace81,Wormer94,Schatz98}.
Following the breakout of the hot-CNO cycle the reaction flow in the $sd$-shell up to ${}^{40}\textrm{Ca}$ is determined by a competition between the $(\alpha,p)$ process \citep{Wallace81}, which is very temperature dependent, because of the strong Coloumb-barrier against alpha-capture, and the $rp$-process. 
The latter is only temperature dependent in the sense that the temperature determines the ratio of the proton-rich isotopes within an isotone chain and thus which isotopes $\beta^+$-decay and move the flow between the isotone chains \citep{Rembges97}. 
It is the detailed behavior of this flow which determines the burst behavior following the first peak. 
If the reaction flow is stopped at a ``waiting-point'' isotope, the nuclear energy release rate decreases and causes a dip in the surface luminosity.
The surface luminosity rises again, when the nuclear energy release rate -- following the waiting -- increases again.

\begin{figure}[ht]
 \plotone{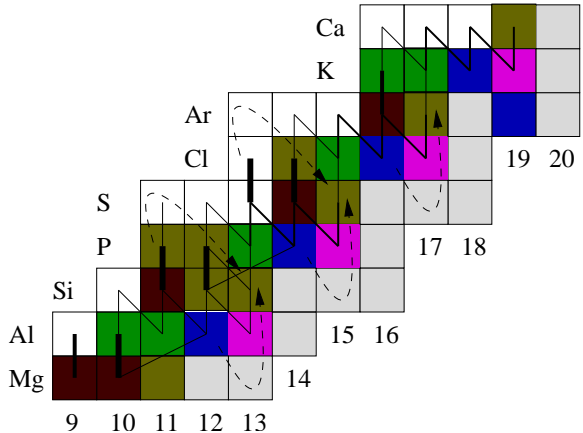} 
 \caption{Targets with $(p,\gamma)$ $Q$-values of $\leq 0.0$ (white), 0.0--0.5 (red), 0.5--2.5 (yellow), 2.5--3.5 (green), 3.5--5.5 (blue), 5.5--8.0 and (purple). Stable nuclei are color-coded grey. The reaction flow rates, $f_{ij}= \dot{Y}_{i\to j} - \dot{Y}_{j\to i}$, during the burst peak temperature are shown in solid lines. The thickness indicates the strength of the reaction flow with the exception of isotopes in $(p,\gamma)$-$(\gamma,p)$ equilibrium which are shown in thick lines though the net flow is close to zero. The reaction paths which circumvent the waiting points are indicated with dashed arrows.}\label{qvalueflow}
\end{figure}

The $\beta^+$-unstable even-even $T_z=-1$ isotopes, ${}^{22}\textrm{Mg}$, ${}^{26}\textrm{Si}$, ${}^{30}\textrm{S}$, and ${}^{34}\textrm{Ar}$, are potential waiting points, which are important, because most of the reaction flow passes through them. 
The reason is that increasing temperatures shift the flow towards the proton dripline, whence only two possible flow paths circumventing them need consideration (see fig.~\ref{qvalueflow}). 
The first path depends on branching between the $\beta^+$-decay ($T_{1/2} \sim \textrm{few seconds}$) of the even-$N$-odd-$Z$ $T_z=(N-Z)/2=-1/2$ isotope and the proton-capture going to the potential waiting point isotope, but because of the high temperature the proton-capture rate is very much larger, which makes the decay insignificant in comparisson.
The second path would depend on double $\beta^+$-decays of very proton-rich isotopes between successive odd-Z isotones. 
These decays proceed through the even-$Z$ isotones where the $(p,\gamma)$-$(\gamma,p)$ reactions generally shift the equilibrium towards the waiting points.

Since ${}^{22}\textrm{Mg}$, ${}^{30}\textrm{S}$, and ${}^{34}\textrm{Ar}$ have low $(p,\gamma)$ $Q$-values (colored red on fig.~\ref{qvalueflow}) they will enter $(p,\gamma)$-$(\gamma,p)$ equilibrium with  ${}^{23}\textrm{Al}$, ${}^{27}\textrm{P}$, ${}^{31}\textrm{S}$, and ${}^{35}\textrm{K}$ respectively \citep{Thielemann94}, whence the ratio of the target and the compound isotope is only determined by a Boltzmann factor, $\exp(-k_BT/Q)$, along with a spin statistics factor.
For ${}^{22}\textrm{Mg}$ ($Q=126\textrm{ keV}$), ${}^{30}\textrm{S}$ ($Q=296\textrm{ keV}$), and ${}^{34}\textrm{Ar}$ ($Q=78\textrm{ keV}$) the $Q$-values are sufficiently low to shift the ratio towards the for photo-disintegration to prevent a significant flow through the proton-capture compound isotope.
This is not the case for ${}^{26}\textrm{Si}$ ($Q=859\textrm{keV}$) where the equilibrium balance is shifted towards ${}^{27}\textrm{P}$, so unless the $Q$-value has been overestimated, ${}^{26}\textrm{Si}$ is not a waiting point, because it is destroyed via  the $\beta^+$-decay of ${}^{27}\textrm{P}$.

In addition to the $(p,\gamma)$-$(\gamma,p)$ equilibrium and the $\beta^+$-decay a third branching of the flow from these isotopes originates in the $(\alpha,p)$-process which commences on ${}^{14}\textrm{O}$ and proceeds via ${}^{14}\textrm{O}(a,p)(p,\gamma)$ ${}^{18}\textrm{Ne}(a,p)(p,\gamma)$ ${}^{22}\textrm{Mg}(a,p)(p,\gamma)$ ${}^{26}\textrm{Si}(a,p)(p,\gamma)$ ${}^{30}\textrm{S}(a,p)(p,\gamma)$ ${}^{34}\textrm{Ar}(a,p)(p,\gamma)$ ${}^{38}\textrm{K}$.

\begin{figure}[ht]
\plotone{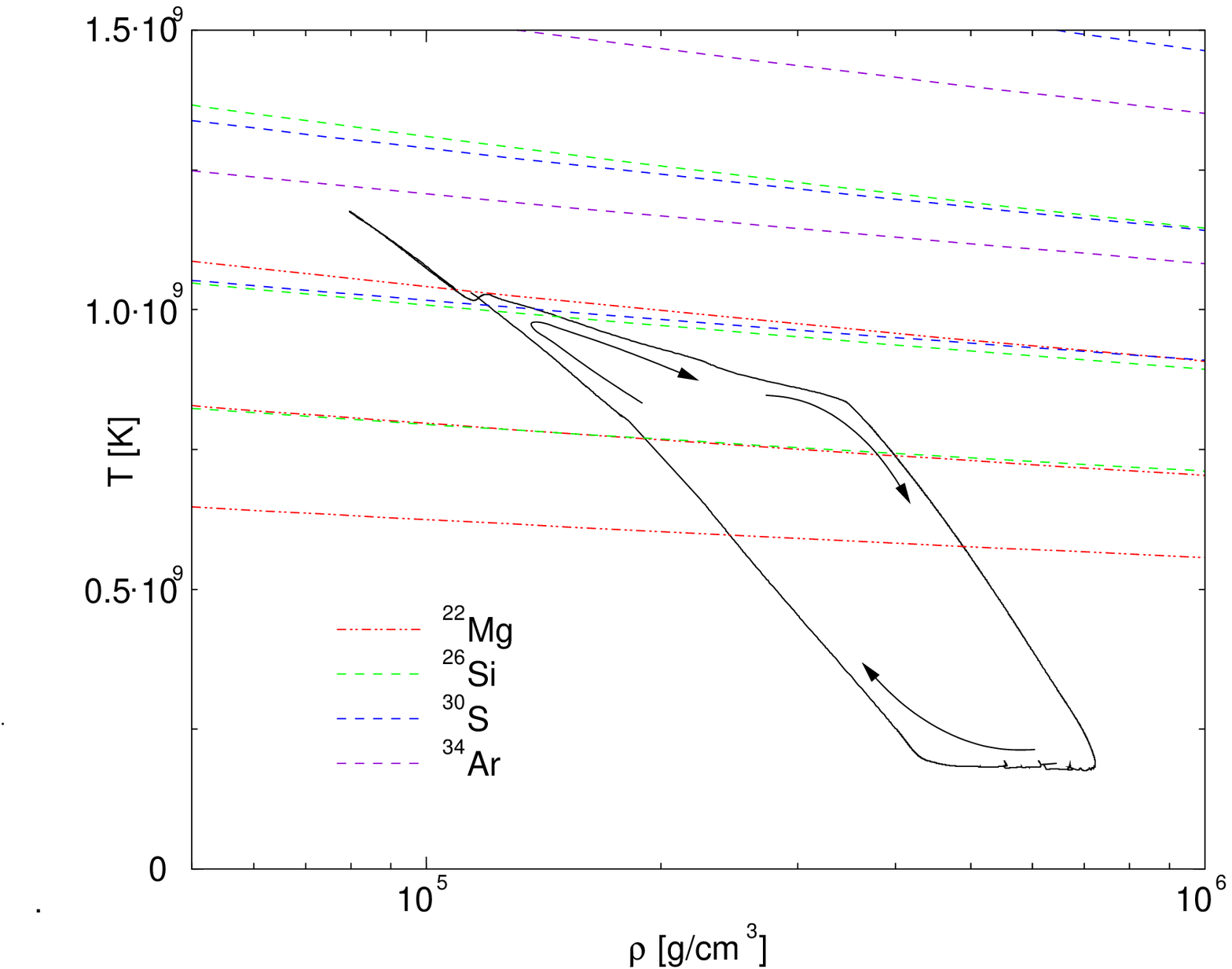}
\caption{The demarcation lines of eq.~[\ref{eq:betaap}] are plotted thrice with $\Delta(T)=\{10^{-2},1,10^2\}$ and $Y_\alpha=0.125$ for the ${}^{22}\textrm{Mg}$, ${}^{26}\textrm{Si}$, ${}^{30}\textrm{S}$, and ${}^{34}\textrm{Ar}$ isotopes. The demarcation line for underestimating ${}^{34}\textrm{Ar}(\alpha,p){}^{37}\textrm{K}$ by a factor $10^2$ falls outside the graph. In addition the solid line shows the thermodynamic trace of the shell conditions during a full burst cycle which is computed in section \ref{computation}. The arrows indicate the time direction of the cycle.}\label{56d0alpha}
\end{figure}

An isotope is a potential waiting points if the $(\alpha,p)$-reaction is so weak that the major part of flow must await $\beta^+$-decay. 
Fig.~\ref{56d0alpha} shows a plot of the condition
\begin{equation}\label{eq:betaap}
\lambda_{\beta^+} = N_A Y_\alpha \Delta(T) <\sigma v>_{(\alpha,p)}\,,
\end{equation}
which demarcates the $(\rho,T)$-space between a dominant $\beta^+$-decay (below the line) and a dominant proton capture (above the line) assuming that $Y_\alpha=0.5/4=0.125$. 
The plot ignores the $(p,\gamma)$-reaction leak which would move the line downwards.
The temperature dependent function $\Delta(T)$ accounts for the uncertainty in the reaction rate which in this paper is set to be constant for the sake of simplicity.
Plotting eq.~[\ref{eq:betaap}] with $\Delta(T)=10^2$ and $\Delta(T)=10^{-2}$ gives an estimate of the effect of the uncertainty of the reaction rate. 

From fig.~\ref{56d0alpha} it is seen that ${}^{22}\textrm{Mg}(\alpha,p){}^{25}\textrm{Al}$ dominates ${}^{22}\textrm{Mg}(\beta^+){}^{22}\textrm{Na}$ even if ${}^{22}\textrm{Mg}(\alpha,p){}^{25}\textrm{Al}$ had been underestimated by a factor 100. It is therefore unlikely that ${}^{22}\textrm{Mg}$ is a waiting point.
However, both ${}^{30}\textrm{S}(\alpha,p){}^{33}\textrm{Cl}$ and ${}^{34}\textrm{Ar}(\alpha,p){}^{37}\textrm{K}$ block the flow during the burst conditions, because the temperature never rises sufficiently to break the Coulomb barrier.

Therefore the destruction rate (see e.g.~\cite{Wormer94}) of the waiting point isotopes depends on the interplay between three rates whose exact details depend on nuclear data and the density and temperature through the interaction with the hydrodynamics. 

\section{Computational results and conclusion}\label{computation}
The combined effect of the two peaks has been computed using a new self-consistent one-dimensional model. It is based on modified version of \verb+AGILE+ \citep{Liebendoerfer02} with added turbulent mixing using a diffusion equation and convective heat transport based on Mixing Length Theory. 
The code tracks energy transport to a high precision and realistically considers the heat transport between the atmosphere and the core.
The reaction network, which includes proton-, neutron-, and alpha-induced reactions and accounts for neutrino losses, comprises 298 isotopes extending up to Tellurium. 
Further technical details will be described in an upcoming paper \citep{Fisker03d}.

\begin{figure}[ht]
 \plotone{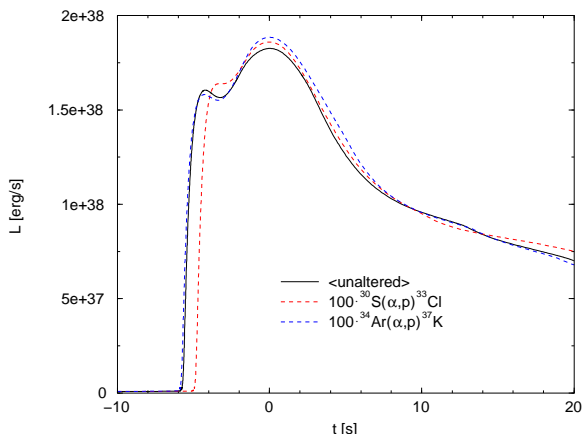} 
 \caption{The computed luminosity is plotted as a function of time for the three cases. The three luminosity curves have been synchronized to make the second burst peaks coincide.}\label{56dLt}
\end{figure}

Having eliminated ${}^{22}\textrm{Mg}$ and ${}^{26}\textrm{Si}$, we investigate ${}^{30}\textrm{S}$ and ${}^{34}\textrm{Ar}$ by assuming a local rest mass accretion rate of $\dot{M}=5\cdot10^{16}\textrm{g}/\textrm{s}$ and computing the bolometric luminosity curves for the current reaction-rates which are based on the code \verb+NON-SMOKER+ \citep{Rauscher01}. 
In addition luminosity curves were computed assuming that the ${}^{30}\textrm{S}(\alpha,p){}^{33}\textrm{Cl}$ and ${}^{34}\textrm{Ar}(\alpha,p){}^{37}\textrm{K}$-reaction rates have been underestimated by a factor 100 i.e.~$\Delta(T)=100$. 
This implicitly tests any uncertainty in the $\beta+$-decay half lives.
The luminosity curves are shown in fig.~\ref{56dLt}, while a trace of the temperature and density is shown in fig.~\ref{56d0alpha} using the unaltered reaction rates.
Fig.~\ref{56dLt} shows that if the ${}^{30}\textrm{S}(\alpha,p){}^{33}\textrm{Cl}$-reaction were faster then the dip in the luminosity would decreased or possibly be eliminated. 
Increasing the ${}^{34}\textrm{Ar}(\alpha,p){}^{37}\textrm{K}$-reaction decreases the dip slightly indicating that the ${}^{34}\textrm{Ar}$-waiting point plays a smaller role in creating the full dip in the unaltered model. 
In both cases increasing the rate of either $(\alpha,p)$-reaction increases the luminosity of the second peak indicating that a combination would increase further.

Convective heat transport defines the rise of the first burst peak, whence a time dependent model of the turbulent part of the kinetic energy could change the shape or magnitude of the first burst peak.
In addition a more effective convective heat would increase the temperature in the upper regions which would make the $(\alpha,p)$-reaction flow stronger despite the resulting drop in density from the temperature increase and vice versa.

In conclusion the most uncertain factors for the impedance driven double peak is the model of convection, the accretion rate, the restriction to a spherically symmetric model and the uncertainty in the theoretical $(\alpha,p)$ rates ${}^{30}\textrm{S}$ and ${}^{34}\textrm{Ar}$. 
However, even if the effect may not be decidedly observable in the light curve, the waiting point impedance of ${}^{30}\textrm{S}$ and ${}^{34}\textrm{Ar}$ and possibly ${}^{22}\textrm{Mg}$ and ${}^{26}\textrm{Si}$ is inherent to the system and an experimental confirmation of the theoretical rates would help to determine the limits of the fast flow through the $sd$-shell isotopes, whereas improved hydrodynamics is important to affirm the $(\rho,T)$-parameter space of the  $(\alpha,p)$-reactions.

\acknowledgements
JLF and FKT acknowledge support from the Swiss NSF grant 2000-061822.

\end{document}